\documentclass[aps,twocolumn,groupedaddress,showpacs,floatfix]{revtex4}

\usepackage{graphicx}

\begin{document}

\title{Influence of the Radio-Frequency source properties on RF-based atom traps}
\author{O.~Morizot}
\author{L.~Longchambon}
\author{R.~Kollengode Easwaran}
\author{R.~Dubessy}
\author{E.~Knyazchyan}
\author{P.-E.~Pottie}
\author{V.~ Lorent}
\author{H.~Perrin}\email{helene dot perrin at univ - paris13 dot fr}

\affiliation{Laboratoire de physique des lasers, Institut Galil\'ee, Universit\'e Paris 13 and CNRS,
Avenue J.-B. Cl\'ement, F-93430 Villetaneuse, France }

\date{\today}

\begin{abstract}
We discuss the quality required for the RF source used to trap neutral atoms in RF-dressed potentials. We illustrate this discussion with experimental results obtained on a Bose-Einstein condensation experiment with different RF sources.
\end{abstract}

\pacs{39.25.+k, 32.80.Pj}
%
%

\maketitle

\section{Introduction}
Radio-frequency (RF) fields are used in cold atom experiments for different purposes: for instance, evaporative cooling performed in a magnetic trap relies on RF field coupling between the different atomic magnetic states~\cite{Hess1986,Hess1988}. This technique led to the first observation of Bose-Einstein condensation (BEC)~\cite{Cornell2002,Ketterle2002}. RF pulses are used for dissociating ultracold molecules produced from ultracold gases through Feshbach resonances~\cite{Regal2003}. More recently, RF fields have been used together with static magnetic fields for trapping ultracold atoms at a temperature of a few $\mu$K in unusual geometries~\cite{Zobay2001,Colombe2004}. There is a growing interest for these ``RF-based traps'' among atomic physicists, for creating double well traps on atom chips~\cite{Schumm2005,Extavour2006,Jo2007} as well as proposing new kinds of confining potentials~\cite{Lesanovsky2006,Courteille2006,Morizot2006,Lesanovsky2007}. In both cases, a single frequency RF signal must be frequency swept over some range, often larger than the initial frequency, following a precise time function lasting several seconds. Typically the RF frequency is varied between 1\,MHz and a few tens of MHz in 0.1\,s to 10\,s in a first ramping stage, and held at the final frequency for seconds in a plateau stage. Commercial RF generators meet reasonably well the usual requirements for evaporative cooling, even if a better resolution in arbitrary frequency ramps would be appreciated. However, in the case of RF-based trapping, the requirements are stronger. The main difference between these two situations is that in evaporative cooling the cold atomic sample is located away from the region of efficient coupling, whereas in the RF-based trapping scheme the coldest atoms sit exactly at the point where the RF field has the largest effect. The quality of the RF source is then much more important than for evaporative cooling. Indeed, the cloud position is directly related to the value of the RF field frequency, whereas the trap restoring force, or equivalently the oscillation frequency $\nu_t$ in the harmonic approximation, is linked to the RF amplitude. As a result, any amplitude noise, frequency noise or phase noise of the RF signal during the ramp or the plateau leads to heating of the cold atomic cloud~\cite{Colombe2004,White2006}. These considerations motivated the detailed study presented in this paper.

This paper is organised as follows: In section~\ref{Sec:Theory} we give explicit expressions for the heating of the cold atom sample for frequency and amplitude noise in the case of RF-based trapping. Section~\ref{Sec:Results} is devoted to experimental results, with a comparison between different RF sources tested on the BEC experiment.

\section{Requirements on the RF source for RF-based trapping}
\label{Sec:Theory}
In this paper, we will focus on the RF-dressed trap that is experimentally produced in our laboratory~\cite{Colombe2004}. The main conclusions may easily be extended to other RF-dressed trap geometries.

The trap confines the atoms in all three space dimensions. The trapping force arises from the interaction between the atoms and the linearly polarized RF field $B(t) = B_{\rm RF} \cos(2\pi \nu_{\rm RF} t)$ in the presence of an inhomogeneous static magnetic field $B_{\rm dc}$. This interaction dresses the atoms with RF photons and results in a transverse confinement of the dressed atoms to an isomagnetic surface. The atoms are free to move along this confining surface, resulting in our case in a kind of ``bubble trap''~\cite{Zobay2001}. Due to gravity, however, they are concentrated at the bottom of the surface. Their motion is pendulum-like in the horizontal directions, and imposed by the RF interaction along the vertical $z$ axis. This last direction is thus the most sensitive to the RF field properties (frequency $\nu_{\rm RF}$, amplitude $B_{\rm RF}$) and we will therefore concentrate on the vertical motion in the following. Along this direction, heating or atomic losses may arise from frequency or amplitude noise, phase jumps or discrete frequency steps during the RF ramp.

\subsection{Frequency noise: Dipolar excitation heating}

Very generally, for atoms in a one dimensional harmonic trap with a trapping frequency $\nu_z$, any effect producing a jitter in the trap position $z$ results in linear heating through dipolar excitation. The average energy of the cold atomic cloud $E$ increases linearly as~\cite{Gehm1998}:
\begin{equation}
\dot{E}=\frac{1}{4} M \omega_z^4 \, S_{z}(\nu_z) \label{eqn:dipolar_heating}
\end{equation}
where $\omega_z = 2 \pi \nu_z$, $M$ is the atomic mass and $S_{z}$ is the one-sided Power Spectral Density (PSD) of the
position fluctuations $\delta z$, defined as the Fourier transform of the time correlation function~\cite{Gehm1998}
\begin{equation}
S_z(\nu) = 4 \int_0^{\infty} \! \! d\tau \, \cos(2 \pi \nu \tau) \langle \delta z(t) \, \delta
z(t+ \tau) \rangle. \label{eqn:Sz}
\end{equation}
In our 3D trap, the time variations of energy, $E$, and temperature, $T$, are related by $\dot{T}=\dot{E}/3k_B$. In principle, $\dot{E}$ is the sum of three contributions of the type~(\ref{eqn:dipolar_heating}) for all three directions. However, the vertical heating is always much larger than the horizontal ones in this RF-based trap, and we will neglect the minor contributions in the following.

The vertical trap position $z$ is a function $z=Z(\nu_{\rm RF})$ of the RF frequency $\nu_{\rm RF}$. As a result,  $S_{z}$ is directly proportional to $S_{\rm rel}$, the PSD of relative frequency noise of the RF source:
\begin{equation}
S_z(\nu) = \left(\nu_{\rm RF} \frac{d Z}{d \nu_{\rm RF}} \right)^2 S_{\rm rel}(\nu).
\label{eqn:Sz_to_Sy}
\end{equation}
The function $Z$ depends on the geometry of the static magnetic field. In a quadrupolar field, for instance, $Z$ is linear with $\nu_{\rm RF}$ and its derivative is a constant. From Eqs.~(\ref{eqn:dipolar_heating}) and (\ref{eqn:Sz_to_Sy}), we infer that the linear heating rate is proportional to $S_{\rm rel}(\nu_z)$.

To fix orders of magnitude, within the static magnetic field of our Ioffe-Pritchard trap~\cite{Colombe2004}, $\nu_z$ may be adjusted between 600 and 1500\,Hz and the typical temperature of the cold rubidium 87 atoms ranges from 0.5 to 5\,$\mu$K. For Bose-Einstein condensation experiments, a linear temperature increase below 0.1\,$\mu$K.s$^{-1}$ is desirable~\cite{Ketterle1999}. This rate corresponds to $\sqrt{S_{z}(\nu_z)} = 0.3$\,nm/$\sqrt{\rm Hz}$ for an intermediate trap frequency of 1000\,Hz and $\nu_{RF} = 3$\,MHz, which in turn corresponds to a one-sided PSD of relative frequency fluctuations of the RF source $S_{\rm rel}(\nu_z) = -118$\,dB.Hz$^{-1}$.

\subsection{Amplitude noise: Parametric heating}
Fluctuations of the RF field amplitude $B_{\rm RF}$ are responsible for parametric heating in the vertical direction. The trapping frequency $\nu_z$ is inversely proportional to $\sqrt{B_{\rm RF}}$~\cite{Zobay2001}:
\begin{equation}
\nu_z = \left(\frac{dZ}{d\nu_{\rm RF}}\right)^{-1}\sqrt{\frac{2F\hbar}{M\gamma B_{\rm RF}}} .
\end{equation}
Here, $\gamma$ is the gyromagnetic ratio of the atom and $F$ is the total atomic spin ($F=2$ for rubidium 87 in its upper hyperfine state). The atoms are assumed to be polarised in their extreme $m_F=F$ substate. Amplitude noise then results in an exponential increase of the cloud temperature with a rate $\Gamma$ given by
\begin{equation}
\Gamma = \pi^2 \nu_z^2 S_a(2 \nu_z)
\end{equation}
where $S_a$ is the PSD of the relative RF amplitude noise~\cite{Gehm1998}. In order to perform experiments with the BEC within a time scale of a few seconds, $\Gamma$ should not exceed $10^{-2}$\,s$^{-1}$ \cite{Ketterle1999}. Again, for a typical oscillation frequency of 1000\,Hz, this corresponds to $S_a < -90$\,dB.Hz$^{-1}$. This requirement is rather easy to match and does not limit the choice of the RF source, as -110\,dB.Hz$^{-1}$ is commonly reached with commercial synthesizers. However, particular care must be taken in the choice and installation of the RF amplifier usually used after the source.

\subsection{Phase jumps}
\label{Sec:PhaseJumps}
Controlling the phase of the RF source is not a crucial point for evaporative cooling, but becomes an issue in the case of RF-based traps, where it is associated with trap losses. In the latter situation, the atomic spin follows an effective magnetic field precessing at the RF frequency around the dc magnetic field, with a precession angle $2\pi\nu_{\rm RF}t$ and a nutation angle $\theta$. A phase jump results in a sudden change $\Delta\varphi$ in the precession angle, the atomic spin being then misaligned with the new direction the effective field. Some of the atoms end up with a spin oriented incorrectly and escape the trap.

The atomic loss after a phase jump $\Delta\varphi$ may be estimated in the following way. The nutation angle $\theta$ is linked to RF frequency and amplitude through $\tan(\theta)= (B_{\rm RF}/2)/(B_{\rm dc} - 2\pi\hbar\nu_{\rm RF}/g_F\mu_B)$~\cite{note1}. $g_F$ and $\mu_B$ are the Land\'e factor and the Bohr magneton, respectively. $\theta$ is position dependent, as is the value of the dc magnetic field $B_{\rm dc}$. After the phase jump, the new effective field makes an angle $\psi$ with the former one, where $\sin(\psi/2) = \sin(\theta) \sin (\Delta\varphi/2)$. At a given position, the probability $p(\psi)$ of keeping an atom  in the new dressed state is then given by its overlap with the initial spin eigenstate, $p(\psi)=[\cos(\psi/2)]^{4F}$. In the case of rubidium 87 in the $F=2$ hyperfine ground state, starting polarized in the $m_F=2$ dressed state, $p(\psi)=\cos^8(\psi/2)$. The fraction  $P(\Delta\varphi)$ of the atoms remaining in the right state is then an average of $\langle p(\psi)\rangle_{\theta}$ over the cloud size. For example, a phase jump of $10^{-2}$\,rad will result at most in a loss of a fraction $10^{-4}$ of the atoms in states other than $m_F=2$.

For a single phase jump, as illustrated experimentally in section~\ref{Sec:Results}, the phase jump amplitude should then remain below 0.1\,rad for limiting the losses to 1\%. However, even much smaller phase jumps should be avoided, if they are repeated. This is difficult to achieve with an analog synthesizer over a wide frequency sweep. By contrast, Direct Digital Synthesis (DDS) technology is well adapted to this requirement\,\cite{DDSBook}.

\subsection{Frequency steps}
\label{Sec:FrequencySteps}
The drawback of DDS technology is that, although the phase is continuous, the frequency is increased in successive discrete steps $\delta\nu$. A sudden change in the RF frequency also results in atomic losses, through the same mechanism as for phase jumps. The effective magnetic field rotates, at most, by the small angle $\delta\theta$ given by $\delta\theta = 2\pi\,\delta\nu/(\gamma B_{\rm RF}/2)$. For a linear ramp with $N$ steps over a frequency range $\Delta\nu = N \delta\nu$, the  fraction of atoms remaining after the full ramp is of order $\left[\cos\left(\delta\theta/2\right)\right]^{4FN}$. Given the expression for $\delta\theta$, this reads:
\begin{equation}\label{Eq:FreqStepCriteria}
\left[\cos\left(\frac{2\pi\,\Delta\nu}{N\gamma B_{\rm RF}}\right)\right]^{4FN} \simeq 1 -
\frac{F}{2N} \left(\frac{4\pi\,\Delta\nu}{\gamma B_{\rm RF}}\right)^2 \, .
\label{eq:fresqteps}
\end{equation}
Thus, for the remaining fraction to be larger than 95\%, the number of frequency steps should be larger than $N_{\rm min}$, where $N_{\rm min} = 10 F (4\pi\,\Delta\nu/\gamma B_{\rm RF})^2$. For example, for a 2\,MHz ramp with a typical RF amplitude of 200\,mG, it yields $N_{\rm min} = $16,000.

In addition to this loss effect, a sudden change in the RF frequency results in a sudden shift of the position of the RF-dressed trap.  This may cause dipolar heating of the atoms, especially if this frequency change occurs every trap period. The frequency steps should thus be as small as possible, a few tens Hz to a hundred Hz typically.

\section{Results}
\label{Sec:Results}
\subsection{Experimental set-up}

\begin{figure}[t]
 \begin{center}
 \includegraphics[width=0.92\columnwidth]{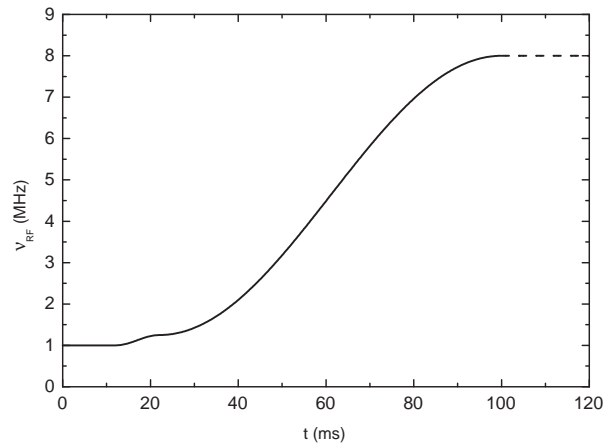}
 \end{center}
\caption{Typical shape of a radio-frequency ramp applied to the ultracold atomic sample. In the present
example $\nu_{RF}$ is increased from 1 to 8\,MHz within 88\,ms, after a 12\,ms stage where the frequency is maintained at 1\,MHz for adiabatic switching of the RF source. At the end of the ramp, the RF frequency is maintained at its final value for some holding time in the RF-based trap, dashed line.}
\label{fig:ramp}
\end{figure}

As mentioned in section~\ref{Sec:Theory}, the RF signal is used for producing a bubble-like trap, where ultracold rubidium atoms are accumulated at the bottom of the surface. The trap is very anisotropic with stronger confinement in the vertical direction~\cite{Colombe2004}. For the RF-dressed trap to be efficiently loaded from a Ioffe-Pritchard static magnetic trap, the RF frequency $\nu_{RF}$ is ramped up from 1\,MHz to a final fixed frequency $\nu_{RF}^{end}$ ranging from 2 to 10\,MHz. The static magnetic field, necessary both for magnetic trapping and RF-induced trapping, is always present. A typical ramp is shown on Fig.~\ref{fig:ramp}. The frequency is ramped more slowly around 1.3\,MHz, corresponding to the resonant frequency at the centre of the magnetic trap where adiabaticity of spin rotation is more difficult to obtain. At the end of this ramp, which may last between 75\,ms and 500\,ms, the RF frequency is held between 0.1 and 10\,s for recording the lifetime and heating rate of the atoms in the RF-based trap.

In order to control the RF amplitude, a programmable RF attenuator Minicircuits ZAS-3 is driven by an analog output channel of a National Instrument PC card PCI-6713. This attenuator adds 5.1\,dB of amplitude noise. At the output of the attenuator, the RF signal is amplified by a HD Communications Corp. class-AB amplifier HD19168. Its gain is 40\,dB and its noise figure is typically +7\,dB according to the manufacturer specifications. This 12 dB total excess noise was still too low for the amplitude noise to have a significant effect on the atoms, and we did not observe any parametric heating in our experiments. The RF field is then produced by a small circular antenna of 3\,cm diameter. The field is linearly polarised and its amplitude $B_{\rm RF}$ may be adjusted between 70 and 700\,mG.

In the following, we present three kinds of measurements performed with different RF sources. First, we investigate the effect of a single phase jump at the end of the ramp on the trapped atom number, for a fixed holding time. Second, the number of frequency steps during the ramp is varied and the final number of atoms is recorded together with the temperature. Finally, the holding time is varied to measure the lifetime and heating rate in the RF-based trap, for three different RF-sources. The atom number and the temperature are measured after a ballistic expansion of 10\,ms by absorption imaging on a CCD camera.

\subsection{Phase jump}

\begin{figure}[t]
 \begin{center}
\includegraphics[width=\columnwidth]{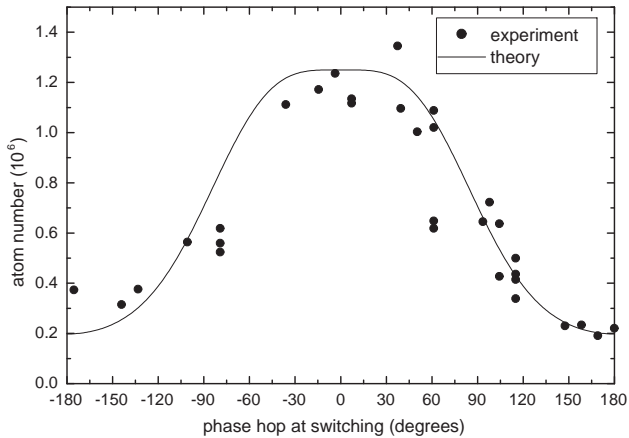}
 \end{center}
\caption{Number of atoms remaining 1\,s after switching between the two synthesizers as a function of the phase jump. Experimental data, full circles, are compared to a calculation, full line, with an RF amplitude of 470\,mG and a temperature of $4\,\mu$K.}
\label{fig:phasejump}
\end{figure}

The effect of a phase jump is tested in the following way. The RF frequency is swept from 1\,MHz to 3\,MHz with a Stanford DS-345 DDS. The 1,500 points frequency ramp is non linear, with a slower zone around the resonance crossing at 1.3\,MHz in the spirit of the typical ramp depicted on Fig.~\ref{fig:ramp}. The RF antenna is then switched to a second independent synthesizer, a Rohde \& Schwartz SML-01 maintained at a fixed frequency of 3\,MHz for the full holding time (dashed line of Fig.~\ref{fig:ramp}). The phase difference is not controlled at the switching, but is monitored for each experiment with an oscilloscope. The final atom number is recorded after a 1\,s holding time. The results are presented on Fig.~\ref{fig:phasejump}, full circles. For the maximum phase jump, $\pi$, $80\%$ of the atoms are lost. This figure depends on the atomic temperature, the losses being higher at lower temperature, and is well reproduced by theory, as shown on Fig.~\ref{fig:phasejump}, black line. The theoretical curve is calculated for the experimental RF amplitude of 470\,mG by averaging the loss probability over the positions of the atoms, as deduced from a thermal distribution at the measured temperature of $4\,\mu$K. The fact that the trap is able to hold two of the five spin components of the $F=2$ hyperfine state is taken into account, by using $p(\psi)=\cos^8(\psi/2) + 4\cos^6(\psi/2)\sin^2(\psi/2)$, see section~\ref{Sec:PhaseJumps} for the definition of the probability $p(\psi)$. The overall amplitude is set to the initial atom number in the magnetic trap before transfer into the RF-based trap, measured independently. We find a very good agreement between the experimental results and the simple theory developed above, with no adjustable parameter.

\subsection{Frequency steps}

As discussed in section~\ref{Sec:FrequencySteps}, the transfer efficiency is expected to depend on the number of frequency steps in the ramp, when a DDS device is used. To evaluate the number of required steps, we repeated the following procedure, varying only the number of frequency points in the arbitrary ramp. The atoms were cooled in the magnetic trap down to $10\,\mu$K. After an adiabatic switching of the RF up to $B_{\rm RF} = 550$\,mG in 12\,ms, the RF frequency was ramped from 1\,MHz to 8\,MHz in 88\,ms to transfer the atoms in the RF-based trap. The number of transferred atoms and their temperature was recorded after 1\,s in the trap. This time was short as compared to the lifetime in the RF-based trap, which reached 25\,s for this experiment. The arbitrary ramp is depicted on Fig.~\ref{fig:ramp}. It consisted in a constant plateau of 12\,ms followed by two connected cosine branches of 10\,ms and 78\,ms respectively, in order to cross the resonance at 1.25\,MHz more slowly. For these experiments, the sine-wave output of a Tabor Electronics DDS synthesizer, model WW1072, was used. The number of frequency points $N$ was varied between 500 and 20,000 which is the maximum value for this device. These frequency lists were constructed by removing uniformly a fraction of the points of a reference arbitrary ramp of 20,000 points. Only 88\% of these points really contributed to the frequency ramp, as 12\% of them were used for maintaining a constant frequency for the initial switching stage.

\begin{figure}[t]
 \begin{center}
\includegraphics[width=\columnwidth]{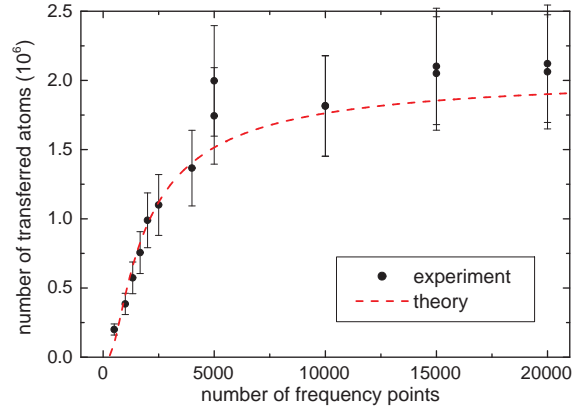}
 \end{center}
\caption{Number of atoms remaining 1\,s after the transfer ramp, as a function of the number of frequency points in the arbitrary ramp. Circles: experimental data. Dashed line: prediction of the model of Eq.(\ref{eq:fresqteps}).}
\label{fig:freqsteptransfer}
\end{figure}

\begin{figure}[t]
 \begin{center}
\includegraphics[width=\columnwidth]{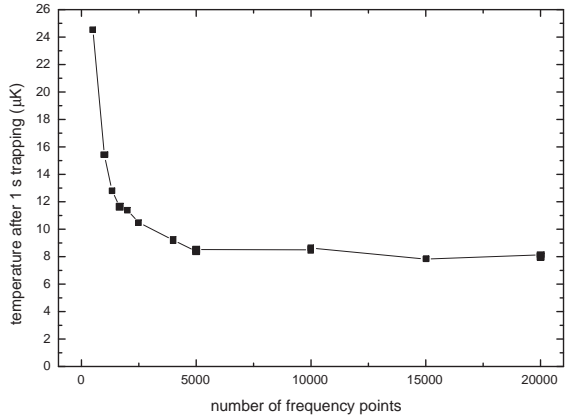}
 \end{center}
\caption{Temperature of the transferred atoms, measured 1\,s after the end of the transfer ramp, as a function of the number of frequency points in the arbitrary ramp. The initial temperature in the magnetic trap was $10\,\mu$K. The solid line is a guide to the eye.}
\label{fig:freqsteptemp}
\end{figure}

The results for atom number and temperature are presented on Fig.~\ref{fig:freqsteptransfer} and \ref{fig:freqsteptemp} respectively. The initial atom number in the magnetic trap was measured to be $2.05\pm 0.2 \times 10^6$. Two regimes are clearly identified. Below $N=$5,000, the number of transferred atoms increases almost linearly with $N$. Above this value, more than 80\% of the atoms are transferred into the RF-based trap, and the number of atoms detected for N larger than 15,000 is even equal to the initial atom number. To compare with the theory of section~\ref{Sec:FrequencySteps}, the expected value of the transferred number given by both the measured initial atom number and Eq.(\ref{eq:fresqteps}) is plotted together with the data, dashed line. The number of frequency steps considered in the calculation is $0.88 N$, as the initial points at constant frequency do not induce losses. The theory reproduces well the overall behaviour, even if it slightly underestimates the transfer efficiency at large $N$. In particular, it predicts a transfer efficiency reaching 80\% for $N=$7,000. The transfer efficiency is expected to reach 93\% for $N=$20,000, which is consistent with the experiment within the uncertainty.

Below 5,000 points, the lack of frequency resolution in the ramp is also visible in the temperature data, Fig.~\ref{fig:freqsteptemp}, as a clear temperature increase during the loading phase. This is explained by the excitation of the dipolar mode through sudden changes in the trap position at each frequency change, as explained in section~\ref{Sec:FrequencySteps}. For more than 5,000 frequency points, the temperature reaches the asymptotic value of $8\,\mu$K. This value is smaller than the initial temperature in the magnetic trap. This was expected as the geometric mean of the oscillation frequencies is much larger in the initial trap. The transfer within 88\,ms is not fully adiabatic in this respect, the expected temperature in the RF-based trap being on the order of $2\,\mu$K for an adiabatic transfer.

\subsection{Frequency noise}

\begin{figure}[t]
 \begin{center}
\includegraphics[width=\columnwidth]{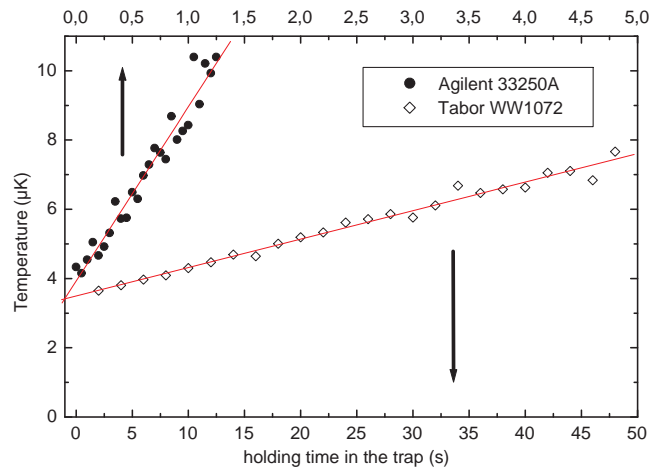}
 \end{center}
\caption{Comparison of heating of the atomic cloud in the bubble trap: Agilent 33250A synthesizer driven by NI board, full circles, or Tabor Electronics WW1072, open diamonds, is used for producing the RF ramp and the
final radio-frequency $\nu_{RF}^{end}$. Note that the horizontal scales differ by a factor 10 for the two data sets. We observe a heating rate of 5.0\,$\mu$K.s$^{-1}$ in the first case and 0.082\,$\mu$K.s$^{-1}$ the second one, as given by a linear fit, full lines. The lifetime reaches 32\,s in this situation.} \label{fig:comparaisonHeating}
\end{figure}

The heating rate in the dressed trap was measured with three different RF sources. First, we used an Agilent 33250A analog synthesizer with an RF amplitude of 500\,mG for both the frequency ramp and the final holding frequency. Such RF analog synthesizers operated at fixed frequency exhibit very good relative frequency noise in most cases, typically at the $-180$\,dB.Hz$^{-1}$ level or better. However, as mentioned by Colombe \emph{et al.}~\cite{Colombe2004} and confirmed by White \emph{et al.}~\cite{White2006}, the relative frequency noise increases by a few decades when a large range frequency sweep is required, as the output frequency then needs to be driven with an external analog voltage. The external voltage control was provided by a PC analog board (NI 6713),  such that the modulation depth was $\pm$1\,MHz on a central frequency of 2\,MHz. We obtained both a short lifetime, typically 400\,ms at $1/e$, and a strong linear heating, as shown on Fig.~\ref{fig:comparaisonHeating} full circles. The heating rate is measured to be $5.0\,\mu$K.s$^{-1}$. This rate, given the RF amplitude, corresponds to a relative frequency noise of $S_{\rm rel}=-100$\,dB.Hz$^{-1}$ at the trap frequency of 600\,Hz, see section~\ref{Sec:Theory}. An independent measurement of the spectral width of the RF signal produced by the Agilent synthesizer in the same conditions indeed gave the same value for the relative frequency noise $S_{\rm rel}=-100$\,dB.Hz$^{-1}$. This noise is quite high because the frequency is varied with a large modulation depth ($\Delta f/f = 1$) and the voltage noise of the NI board is directly translated into frequency noise.

We also measured the lifetime and heating rate with the setup used for phase jump characterisation, with the disadvantage however that the random phase jumps resulted in a large dispersion in the atom number data. Nevertheless, within a precision limited to 0.1\,$\mu$K.s$^{-1}$ and 2\,s respectively, we observed no heating and a lifetime of order 4.5\,s~\cite{Colombe2004}. These good results are linked to the excellent frequency stability of the second device used at fixed frequency.

Finally, the lifetime and heating measurement was repeated with the Tabor Electronics WW1072 device mentioned previously. The loading ramp was the same as described above, with 20,000 frequency points between 1 and 8\,MHz, but with a total duration of 500\,ms. Typical results are presented on Fig.~\ref{fig:comparaisonHeating}, open diamonds. The contrast with the Agilent data is very strong. With the Tabor device and an RF amplitude of 550\,mG, the lifetime increased up to 32\,s, a value comparable with the lifetime in the static magnetic trap. A small linear heating rate of $82\pm5$\,nK.s$^{-1}$ is still present in the RF-based trap for this data set. No exponential parametric heating is measurable. The residual heating rate is slightly larger than the one predicted from the Tabor device specifications. At 8\,MHz and for an oscillation frequency of 600\,Hz, we expect a linear heating rate of 17\,nK.s$^{-1}$. This value is slightly smaller than 33\,nK.s$^{-1}$, which is the lowest observed heating rate with the Tabor device.

\section{Conclusion}
In this paper, we set the requirements on the RF source quality for trapping in RF-induced adiabatic potentials. We present the effects of frequency noise, amplitude noise, phase jumps and finite frequency resolution. The consequence of a frequency jitter, discrete frequency steps and a random phase jump are then demonstrated experimentally.

RF source requirements may be divided into two parts, concerning the ramping stage and the holding stage. Once the atoms are loaded in the RF-based trap, heating and trap losses are avoided if the relative frequency noise and the amplitude noise are below $S_{\rm rel}(\nu_z) = -118$\,dB.Hz$^{-1}$ and $S_a < -90$\,dB.Hz$^{-1}$ respectively. This is relatively easy to meet, although care has to be taken on the choice of low noise RF attenuators and amplifiers. The best results were obtained with the Tabor Electronics device, with a typical heating rate of $0.082\,\mu$K.s$^{-1}$ and a lifetime of up to 32\,s. Finally, the lifetime would be affected if the RF amplitude is below a few tens of mG, as Landau-Zener losses then occur at the avoided crossing~\cite{Zobay2004}.

For the initial frequency sweep, the use of DDS technology ensures phase continuity, and losses or heating are limited if the number of frequency points is large enough, say, 10,000 at least for a few MHz ramp. In the loading stage, once the number of frequency steps is sufficient, the heating will mostly be given by the non adiabatic deformation of the trapping potential when the atoms are transferred from the magnetic to the dressed trap. As the value of the lowest oscillation frequency in the dressed trap is of the order of a few Hz, this heating may be difficult to avoid. On the other hand, cooling directly in the RF-based trap is made possible by the long lifetimes and low heating rates reported here~\cite{Garrido2006,Hofferberth2006}. In any case, these good performances make the RF-based trap compatible with the confinement of Bose-Einstein condensates.

\acknowledgments
This work was supported by the R\'egion Ile-de-France (contract number E1213) and by the European
Community through the Research Training Network `FASTNet' under contract number HPRN-CT-2002-00304 and Marie Curie
Training Network `Atom Chips' under contract number MRTN-CT-2003-505032. Laboratoire de physique des lasers is UMR 7538
of CNRS and Paris 13 University. The LPL is a member of the Institut Francilien de Recherche sur les Atomes Froids.


\begin{thebibliography}{40}
\bibitem{Hess1986} H.~F.~Hess, Phys. Rev. B \textbf{34}, 3476 (1986)
\bibitem{Hess1988} N.~Masuhara, J.~M.~Doyle, J.~C.~Sandberg, D.~Kleppner, T.~J.~Greytak, H.~F.~Hess, and G.~P.~Kochanski, Phys. Rev. Lett. \textbf{61}, 935 (1988)
\bibitem{Cornell2002} E.~A.~Cornell and C.~E.~Wieman, Rev. Mod. Phys. \textbf{74}, 875 (2002)
\bibitem{Ketterle2002} W.~Ketterle, Rev. Mod. Phys. \textbf{74}, 1131 (2002)
\bibitem{Regal2003} C.~A.~Regal, C.~Ticknor, J.~L.~Bohn, and D.~S.~Jin, Nature \textbf{424}, 47 (2003)
\bibitem{Zobay2001} O.~Zobay and B.~M.~Garraway, Phys. Rev. Lett. \textbf{86}, 1195 (2001)
\bibitem{Colombe2004} Y.~Colombe, E.~Knyazchyan, O.~Morizot, B.~Mercier, V.~Lorent, and H.~Perrin, Europhys. Lett. \textbf{67}, 593 (2004)
\bibitem{Schumm2005} T.~Schumm, S.~Hofferberth, L.~M.~Andersson, S.~Wildermuth, S.~Groth, I.~Bar-Joseph, J.~Schmiedmayer, and P.~Kr\"{u}ger, Nature Physics \textbf{1}, 57 (2005)
\bibitem{Extavour2006} M.~H.~T.~Extavour, L.~J.~Le Blanc, T.~Schumm, B.~Cieslak, S.~Myrskog, A.~Stummer, S.~Aubin, and J.~H.~Thywissen, \textit{Proceedings of the International Conference on Atomic Physics}, Atomic Physics \textbf{20}, 241 (2006)
\bibitem{Jo2007} G.-B.~Jo, Y.~Shin, T.~A.~Pasquini, M.~Saba, W.~Ketterle, and D.~E.~Pritchard, M.~Vengalattore, and M.~Prentiss, Phys. Rev. Lett. \textbf{98}, 030407 (2007)
\bibitem{Lesanovsky2006} I.~Lesanovsky, T.~Schumm, S.~Hofferberth, L.~M.~Andersson, P.~Kr\"{u}ger, and J.~Schmiedmayer, Phys. Rev. A \textbf{73}, 033619 (2006)
\bibitem{Courteille2006} Ph.~W.~Courteille, B.~Deh, J.~Fort\`{a}gh, A.~G\"{u}nther, S.~Kraft, C.~Marzok, S.~Slama, and C.~Zimmermann, J. Phys. B: At. Mol. Opt. Phys. \textbf{39}, 1055 (2006)
\bibitem{Morizot2006} O.~Morizot, Y.~Colombe, V.~Lorent, H.~Perrin, and B.~M.~Garraway, Phys. Rev. A \textbf{74}, 023617 (2006)
\bibitem{Lesanovsky2007} I.~Lesanovsky and W.~von Klitzing, Phys. Rev. Lett. \textbf{99}, 083001 (2007)
\bibitem{White2006} M.~White, H.~Gao, M.~Pasienski, and B.~DeMarco, Phys. Rev. A \textbf{74}, 023616 (2006)
\bibitem{Gehm1998} M.E.~Gehm, K.M.~O'Hara, T.A.~Savard, and J.E.~Thomas,  Phys. Rev. A \textbf{58}, 3914 (1998)
\bibitem{Ketterle1999} W.~Ketterle, D.~S.~Durfee, and D.~M.~Stamper-Kurn, \textit{Making, probing and understanding Bose-Einstein condensates}, \textit{in} Proceedings of the International School of Physics ``Enrico Fermi'', Course CXL, edited by M.~Inguscio, S.~Stringari, and C.~E.~Wieman (IOS Press, Amsterdam) p. 67-176 (1999)
\bibitem{note1} The factor $B_{\rm RF}/2$ arises from the fact that only half of the power of the linearly polarized RF has an effect on the atoms.
\bibitem{DDSBook} B.-G.~Goldberg, \textit{Digital Techniques in Frequency Synthesis}, (Mac Graw-Hill, New-York, 1996)
\bibitem{Zobay2004} O. Zobay and B. M. Garraway, Phys. Rev. A \textbf{69}, 023605 (2004)
\bibitem{Garrido2006} C.~L.~Garrido Alzar, H.~Perrin, B.~M.~Garraway, and V.~Lorent, Phys. Rev. A \textbf{74}, {053413} (2006)
\bibitem{Hofferberth2006} S.~Hofferberth, I.~Lesanovsky, B.~Fischer, J.~Verdu and J.~Schmiedmayer, Nature Physics \textbf{2}, {710} (2006)

\end{thebibliography}
\end{document}